# Cerium Oxides without *U*: The Role of Many-Electron Correlation

Tobias Schäfer,* Nathan Daelman,* and Núria López

**ABSTRACT:** Electron transfer with changing occupation in the 4f subshell poses a considerable challenge for quantitative predictions in quantum chemistry. Using the example of cerium oxide, we identify the main deficiencies of common parameter-dependent one-electron approaches, such as density functional theory (DFT) with a Hubbard correction, or hybrid functionals. As a response, we present the first benchmark of ab initio many-electron theory for electron transfer energies and lattice parameters under periodic boundary conditions. We show that the direct random phase approximation clearly outperforms all DFT variations. From this foundation, we, then, systematically improve even further. Periodic second-order Møller−Plesset perturbation theory meanwhile manages to recover standard hybrid functional values. Using these approaches to eliminate parameter bias allows for highly accurate benchmarks of strongly correlated materials, the reliable assessment of various density functionals, and functional fitting via machine-learning.

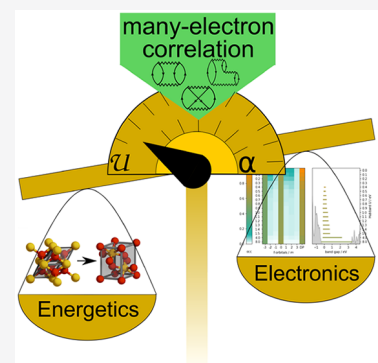

Ab initio techniques at the atomic level hold a key role in materials science and catalysis. DFT in particular is popular since it manages to strike the best balance between accuracy and computational cost. With its widespread success though it is well-documented that (semi)local DFT tends to overly delocalize valence states and wrongly assign them a metallic nature.[1−3] Strongly correlated materials with a high *self-interaction error* are impacted most, including metal oxides, perovskites, and spinels. Ceria, having amassed a strong body of literature because of its industrial importance, illustrates the overall evolution of correlation treatment in multivalency. There has been a clear progression in periodic cell calculations from approximating the localized *f*-electrons by contracting them into the core and neglecting hybridization[4−6] to penalizing off-diagonal occupation via a Hubbard correction *U*,[7,8] the current state-of-the-art. In case higher accuracy is required, hybrid functionals are employed that replace a fraction $\alpha$ of the DFT exchange by the exact Fock exchange.[7,9,10] The latter has become the standard for benchmarking several multivalent material classes, such as perovskites[11] and spinels.[12]

Both options introduce external parameters that affect other observables, for instance the bandgap and electron transfer. The latter plays a key role in ceria chemistry and that of reducible metal oxides at large. Electron-transfer reactions have been linked to polaron hopping,[13] the formation of oxygen vacancies,[14] active oxygen speciation,[15] as well as the dispersion of supported metal nanoparticles.[16−18] Their energy is proportional to the *U* value, both for full[19,20] and partial reduction.[21] The proportionality constant itself remains functional-independent and is, thus, intrinsic to the Hubbard correction.[20] Instead, it changes along the reaction coordinate,[21] introducing a bias either in the thermodynamics, the kinetics, or both. As such, a trade-off has to be made, whereby the optimal *U* may deviate by as much as 1.0 eV in surface reduction.[21] An analogous parameter bias is found for the amount of Fock exchange $\alpha$ in hybrid functionals.[22−24]

In this Letter, we show that systematically improvable quantum many-electron (ME) methods can eliminate these parameter biases in *d*- or *f*-band insulators and semiconductors. We demonstrate this by considering the case study of the electron correlation in the $Ce^{4+} \leftrightarrow Ce^{3+}$ transition as it occurs during the phase transition of bulk $CeO_2$ to $Ce_2O_3$. Moreover, the ME approach also eliminates the parameter-dependence in the lattice constants for both oxides. Being complementary to DFT, they allow for the benchmarking of density functionals[25] and provide high-quality data for constructing or machine-learning optimized density functionals.[26−28] Before, the high computational cost of genuine periodic boundary conditions precluded ME theory for ceria and is only now made possible by reduced complexity algorithms.[29,30]

Specifically, we address two well-known representatives of ME theory: second-order Møller−Plesset perturbation theory (MP2)[31] and the direct particle−hole random phase approximation (RPA).[32,33] We compare them for reference to the Perdew−Burke−Ernzerhof exchange-correlation functional (PBE)[34] with a Dudarev correction, as well as the Heyd−Scuseria−Ernzerhof hybrid functional[35−37] with a

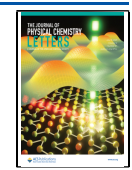

6277

range-separation parameter of 0.3 Å$^{-1}$ (HSE03). Additionally, we demonstrate and discuss the performance of a systematic step beyond the RPA. In ME theories, the electron−electron interaction can be expressed as a sum over one-electron mean-field spin−orbitals

$$\frac{1}{2}\sum_{ij}^{occ}[\langle ij|ij\rangle - \langle ij|ji\rangle] + \frac{1}{2}\sum_{ij}^{occ}\sum_{ab}^{unocc.}\langle ij|ab\rangle[t_{ij}^{ab} - t_{ji}^{ab}] \quad (1)$$

where $t_{ij}^{ab}$ are double excitation amplitudes and the two-electron integrals read

$$\langle pq|rs\rangle = \int d\mathbf{x}\int d\mathbf{x}' \frac{\chi_p^*(\mathbf{x})\chi_r(\mathbf{x})\chi_q^*(\mathbf{x}')\chi_s(\mathbf{x}')}{|\mathbf{r}-\mathbf{r}'|} \quad (2)$$

The double excitation amplitudes $t_{ij}^{ab}$ define the approximation level for the correlation energy (second term in eq 1). The practical computations follow highly optimized formulations and are performed using recent low-scaling MP2[30] and RPA[29] algorithms. The MP2 correlation energy can be defined by $t_{ij}^{ab} = \langle ab|ij\rangle/(\varepsilon_i + \varepsilon_j - \varepsilon_a - \varepsilon_b)$ with Hartree−Fock spin−orbitals $\chi$ and orbital energies $\varepsilon$. The direct RPA correlation energy can be defined by neglecting the exchange-like correlation in eq 1, that is, $t_{ij}^{ab} - t_{ji}^{ab} \to t_{ij}^{ab}$, with amplitudes implicitly defined by

$$t_{ij}^{ab} = [\langle ab|ij\rangle + \langle ak|ic\rangle t_{kj}^{cb} + t_{ik}^{ac}\langle kb|cj\rangle + t_{ik}^{ac}\langle kl|cd\rangle t_{lj}^{db}]$$
$$/(\varepsilon_i + \varepsilon_j - \varepsilon_a - \varepsilon_b) \quad (3)$$

where a sum over occupied $k, l$ and unoccupied $c, d$ spin−orbitals is understood. The solution of eq 3 can analytically be expressed as an infinite sum over all possible ring-like Goldstone diagrams, that is, a summation of solely two-electron integrals and orbital energies, as we illustrate in the Supporting Information (SI). The Kohn−Sham spin−orbitals $\chi$ and orbital energies $\varepsilon$ are provided by the HSE03 functional here.

We model the solid state as primitive bulk unit cells, shown in Figure 1, imposing Born−von Karman periodic boundary conditions and Brillouine zone sampling. Our calculations are carried out using the plane-wave based Vienna ab initio simulation package (VASP).[38] Frozen-core potentials in combination with the projector augmented-wave (PAW)[39] method are used. The cerium atom in particular requires a norm-conserving (nc) PAW potential to ensure proper convergence of the high-energy mean-field orbitals and orbital energies essential for correlation calculations.[40] In combination with extrapolation techniques, the complete basis set and thermodynamic limit of the electron correlation were reached. The SI contains further details and in-depth explanations.

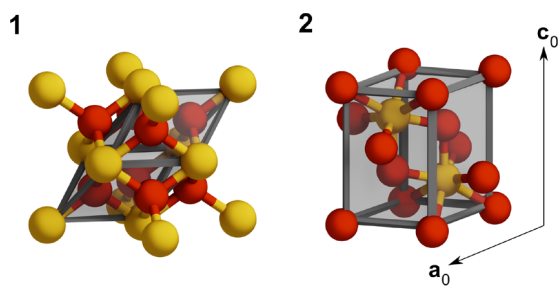

**Figure 1.** Ceria bulk geometries. Panel 1: CeO$_2$ phase exhibiting a fluorite structure (fcc cubic, $Fm\overline{3}m$), of which we show the cubic supercell cell with the primitive cell enclosed in the shaded region. Panel 2: Ce$_2$O$_3$ is a sesquioxide A-type (hexagonal, $P\overline{3}m1$) described by two lattice parameters ($a_0$ and $c_0$). The values of all primitive cell parameters can be found in Table 1. Color code: cerium (yellow), oxygen (red).

To showcase our concerns with the $U$ and $\alpha$ dependence, we performed a sampling for the electronic structure of Ce$_2$O$_3$ in Figure 2. The typical interpretation of Dudarev's implementation is that it localizes the electron density in order to minimize the off-diagonal energy penalty.[41,42] Such a minimum is then reached at integer subshell occupation. In our scan, PBE starts out in a metallic ground state, so that as $U$ increases the more noisy subshells depopulate. Electron density is displaced to outside the $f$-band rather than filling the $f_{z^3} - f_{y(y^2-3x^2)}$ hybridization,[43,44] which mostly remains unaltered. At around $U = 0.4$ eV, a metal-to-insulator transition (MIT) occurs and a bandgap appears. These findings were also recuperated in ref 43. Note that while bulk Ce$_2$O$_3$ passes through a MIT rather early on, defect sites at the CeO$_2$ surface take up to $U = 2.2$ eV for the MIT.[21] Even so, the Hubbard energy penalty remains nonvanishing under the hybridization and instead attains a linear dependence as observed both in this work and others.[7,21] We postulate then that the varying responses at different reaction coordinates in Figure 2 of ref 21 can be traced back to the total filling of the hybridized $f$-orbitals. We further develop a hypothesis for its persistence in the SI.

While the effect of $\alpha$ in HSE03($\alpha$) starts off in a similar fashion to PBE+$U$, the hybridization largely becomes undone in the $\alpha \to 1$ limit. The electron thus starts properly localizing onto $f_{z^3}$, as was the aim of Dudarev's method. This comes at the price though of severely overestimating the O(2p)−Ce 4$f$ gap at 9.45 eV compared to experimental value of 2.4 eV.[45] As such, neither PBE + $U$ nor HSE provide even a qualitatively correct picture of the electronic structure. Recently, Jiang[46] presented with success the first basis set converged, ME correction to the electronic structure of one-electron approaches by using the GW method. By contrast, we target the ground state energy as an integral over the electronic structure. We employ MP2, RPA, as well as corrections beyond RPA, and compare their performance with respect to the one-electron methods on geometry and reaction energy in the following:

First, we evaluate the accuracy of our benchmark with the lattice constants shown in Table 1. Error cancellation is at its greatest here, since we solely compare equivalent condensed phases with only minor variations in the volume or $c/a$ ratio. Deviations from experiments therefore indicate clear, systemic errors in the theory, as can be seen from the PBE + $U$ and HF results for both oxides. The unacceptably large errors of PBE + $U$ exhibit a strong dependence on the $U$ parameter, which is corroborated in refs.[7,23] For both oxides the equilibrium lattice constants $a_0$ and $c_0$ increase with $U$, as illustrated in Figure 2 of ref 7. This has particularly troubling implications for Ce$_2$O$_3$, where $c_0$ grows in the wrong direction and consistently overshoots the experimental value. We arrive at the same conclusion by comparing PBE with PBE + $U$ in this letter. The HF error meanwhile traces back to the missing electron correlation that significantly tightens all chemical bonds. The hybrid functional HSE03 then manages to strike a balanced description at the standard fraction of $\alpha = 0.25$, but presents a strong response in $a_0$ and $c_0$ to the parameter $\alpha$, as was observed in ref 23.

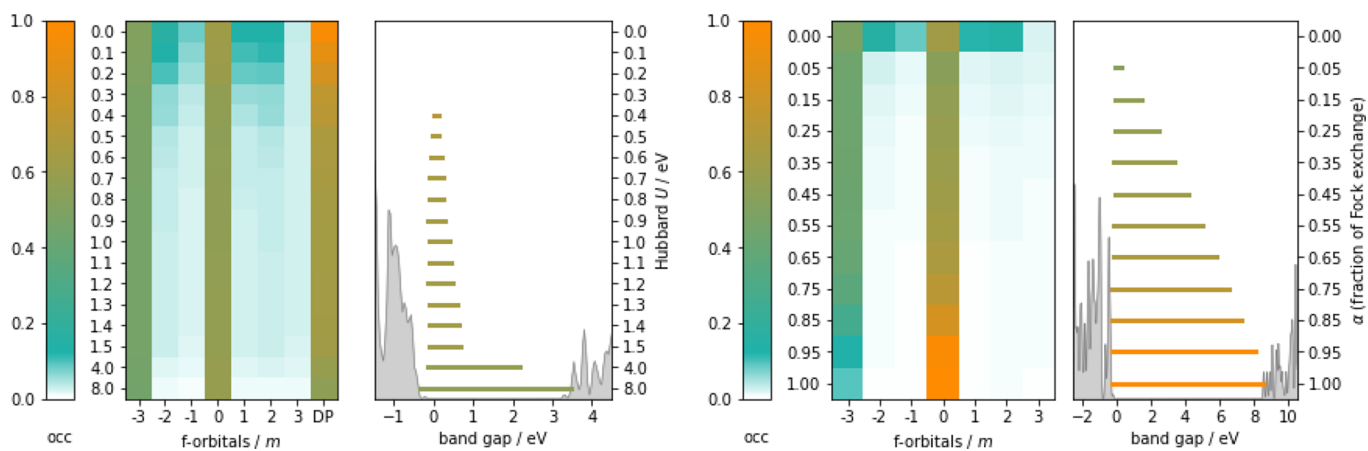

**Figure 2.** Influence of the parameters $U$ for PBE+$U$ (left) and $\alpha$ for HSE03($\alpha$) (right) on the electronic structure of antiferromagnetic $Ce_2O_3$. The heat maps trace the $f$-orbital occupations by magnetic quantum number ($m = -3, ..., +3$) as $U$ and $\alpha$ vary. The Dudarev energy penalty (DP) in the PBE+$U$ heat map measures the orbital-dependent part of the Hubbard correction ($\sum_f n_f(1-n_f)$). The bar plots meanwhile show the matching band gap with at the bottom the accompanying DOS for $U$ = 8.0 eV/$\alpha$ = 1.0.

Table 1. Bulk Properties of $CeO_2$ (Left) and Antiferromagnetic $Ce_2O_3$ (Right)[a]

| | $CeO_2$ | | | $Ce_2O_3$ | | | | | |
|---|---|---|---|---|---|---|---|---|---|
| method | $a_0$ [Å] | % error | $B_0$ [GPa] | $a_0$ [Å] | % error | $c_0$ [Å] | % error | $B_0$ [GPa] | ref |
| RPA | 5.421 | +0.5 | 202 | 3.883 | +0.0 | 6.070 | +0.4 | 145 | |
| RPA+rSOX | 5.377 | −0.3 | 228 | 3.867 | −0.4 | 6.029 | −0.3 | 156 | |
| MP2 | 5.366 | −0.5 | 227 | 3.875 | −0.2 | 6.010 | −0.6 | 145 | |
| dMP2 | 5.381 | −0.2 | 216 | 3.884 | +0.1 | 6.032 | −0.2 | 138 | |
| HF | 5.454 | +1.1 | 229 | 3.929 | +1.2 | 6.259 | +3.5 | 141 | |
| HSE03 ($\alpha$ = 0.25) | 5.399 | +0.1 | 206 | 3.867 | −0.4 | 6.082 | +0.6 | 143 | |
| PBE+$U$ = 4.5 eV | 5.49 | +1.8 | 180 | 3.92 | +1.0 | 6.18 | +2.2 | 111 | 7 |
| PBE | 5.47 | +1.4 | 172 | 3.85 | −0.8 | 6.10 | +0.9 | 101 | 7 |
| LDA+DMFT | | | | 3.81 | −1.9 | | | 164 | 49 |
| expt. | 5.394 | | | 3.882 | | 6.047 | | | 50, 51 |

[a]These include the lattice constants $a_0$ (and $c_0$), the relative error, and the bulk modulus $B_0$. The experimental references are extrapolated to $T \to 0$ K for $CeO_2$ and at $T$ = 3 K for $Ce_2O_3$. Zero-point vibrational effects are not taken into account. In the case of $Ce_2O_3$, the Debye model estimates the experimental lattice constant at 99.7% $a_0$,[47] so that the specified errors are slightly underestimated. For more detail, see the SI.[48]

MP2 greatly improves upon the HF lattice constants for both oxides. The residual underestimation is in fact a general trend of MP2 for solids[52] and goes back to a systemic overestimation of attractive dispersion forces at this level of theory. As a finite-order ME correlation method, the accuracy of MP2 however is limited to the simplest double-excitation effects from the underlying HF Slater determinants. In the language of time-dependent perturbation theory, this in itself excludes electron−hole pair interactions. As such, MP2 cannot adequately account for the correlation effect of electron screening and its quality decreases with increasing polarizability of the material. Note that the static dielectric constant of both cerium oxide phases is quite high, roughly in the order of 25 times the vacuum permittivity.[51,53]

The RPA markedly provides even more accurate lattice constants for $Ce_2O_3$. The remaining overestimation is well-known and may be attributed to the missing exchange-like correlation.[54] The common effect of exchange-like correlation can be well observed when comparing dMP2 (neglecting $-t_{ji}^{ab}$ in eq 1) with MP2 (including $-t_{ji}^{ab}$). By adding exchange-like correlation, the lattice constants are reduced and the bulk modulus grows. A similar effect is observed, when we correct for the exchange-like correlation missing from the RPA. To this end we reintroduce $-t_{ji}^{ab}$ by a renormalized second-order (rSOX) amplitude $-t_{ji}^{ab} = -\langle ab|ji\rangle/(\varepsilon_j + \varepsilon_i - \varepsilon_a - \varepsilon_b - \Delta_{ji}^{ab})$. Here, $\Delta_{ji}^{ab}$ arises from an infinite resummation of Goldstone diagrams. For details on the RPA+rSOX method, we refer to the SI.

Lastly, we address parameter bias in electron transfer reactions, where the total correlation dramatically shifts. In particular, we consider the phase transition from $CeO_2$ to $Ce_2O_3$ under reductive conditions. To prevent issues with the molecular reference (especially $O_2$), we probe the electron transfer energies at three different reactions, labeled $r_1$ to $r_3$. In our notation, $H$ refers to the enthalpy obtained by the VASP calculations. Reaction $r_1$ is catalytically most relevant, as it provides an upper boundary to the enthalpy of oxygen vacancy formation in $CeO_2$. It thus measures the material's reducibility. Hence, the results for reaction $r_1$ are shown in Figure 3, while $r_2$ and $r_3$ are tabulated in the SI.

$$\Delta_{bulk}H_{0K} = H^{Ce_2O_3(s)} - 2H^{CeO_2(s)}$$

$$\Delta_{r_1}H_{0K} = \Delta_{bulk}H_{0K} + \frac{1}{2}H^{O_2(g)}$$

$$\Delta_{r_2}H_{0K} = \Delta_{bulk}H_{0K} + H^{H_2O(g)} - H^{H_2(g)}$$

$$\Delta_{r_3}H_{0K} = \Delta_{bulk}H_{0K} + H^{CO_2(g)} - H^{CO(g)}$$

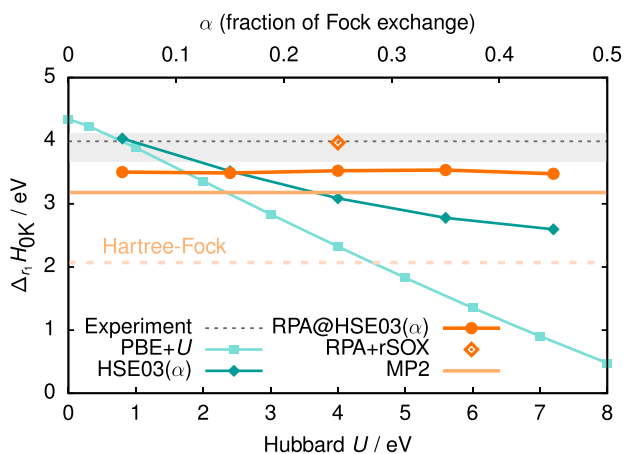

**Figure 3.** Reaction enthalpy, $\Delta_{r_1}H_{0K}$, calculated at different levels of theory and with varying parameters, where appropriate. The Hubbard parameter $U$ is represented by the lower abscissa, the fraction of Fock exchange $\alpha$ by the upper abscissa. The dashed, black line denotes the experimental reference value with an uncertainty indicated by the light gray hatching.[48,55]

Our PBE + $U$ sampling clearly retrieves a linear energy trend, similar to the one found in ref 7. Remarkably, this trend requires Hubbard corrections both on $Ce_2O_3$ but also $CeO_2$, with the latter contributing more to the energy (see SI). In contrast, pristine $CeO_2$ slabs exhibit no $U$ dependence.[21,56] Regardless, such trends are always a source of bias in redox reactions at any specific choice for $U$. If accurate thermodynamics are a priority, however, PBE with a minimal amount of correction already suffices for many transition metal oxides.[7,57−60]

At the same time, there is a clear distinction between parameter-dependent and parameter-independent methods. This is not a trivial observation for our RPA results, which are based on the mean-field orbitals from the HSE03($\alpha$) functional. The variation of the RPA@HSE03($\alpha$) reaction enthalpy is negligibly small (less than 0.06 eV) in the presented range of $\alpha \in [0.05, 0.45]$, but about 1.44 eV for HSE03($\alpha$). Assuming common standard parameter values ($U$ = 4.5 eV[56] and $\alpha$ = 0.25[61]) for PBE + $U$ and HSE03, both MP2 and RPA reach closer agreement to the experimental data. This is very pronounced for the RPA, but less so in the comparison between MP2 and HSE03($\alpha$ = 0.25). In broad terms, the superiority of the ME methods can largely be explained by the systematic nature of the MP2 and RPA errors for atomization energies,[52,58] as can be seen in the SI. The reaction energies, as differences in total atomization energies, thus greatly benefit from systematic error cancellation. This effect is not present in one-electron theories.

Now we examine the different performance of MP2 and RPA in more detail. It is well-known that MP2 reaction enthalpies of small molecules (average error 0.14 eV[62]) exhibit higher quality than reaction enthalpies including extended systems.[63] This reduces the reliability of the mentioned error cancellation between the bulk and the gas in reactions $r_1$−$r_3$. More specifically, we identify a pronounced overestimation of +0.50 eV in the MP2 cohesive energy coming from $Ce_2O_3$.[48] This is strikingly larger than the average systematic error of +0.23[52] in MP2 cohesive energies for solids. The largest error in the MP2 reactions enthalpies can thus be assigned to the $f$-band filling.

The RPA meanwhile constitutes a clear improvement upon MP2. The infinite-order resummation of particle−hole interactions eliminates the technical $\alpha$ dependence of the underlying HSE03 orbitals in the RPA correlation energy. Neither do the dielectric properties cause serious issues for the RPA, as the infinite resummation of ring-diagrams is known to capture electron screening effects with high accuracy.

Despite the RPA outperforming MP2 and all considered DFT flavors, we wish to highlight targets for systematic improvement of the accuracy. While the importance of different classes of Goldstone diagrams missing from the RPA is an ongoing discussion in the literature, we assume the self-correlation error to be particularly critical for the $f$-electrons because of the missing exchange-like correlation. In principle, the RPA self-correlation benefits from error cancellation in energy differences, but the self-correlation of the $f$-bands in $Ce_2O_3$ has no annihilating counterpart in $CeO_2$ nor in the molecules. For comparison, the MP2 exchange term $(-t_{ji}^{ab})$ for instance contributes +0.86 eV to the reaction enthalpy $r_1$, indicating a strong contribution of exchange-like correlation. It is reasonably assumed that this contribution largely corrects the self-correlation error in dMP2. In fact, the simple rSOX correction for the RPA also accounts for a large portion of the self-correlation error, shifting the reaction enthalpy of $r_1$ by +0.45 eV into the experimentally reported region. Although the RPA+rSOX transfer energies for $r_1$−$r_3$ suggest near chemical accuracy, a reliable assessment of the accuracy is impeded by the large uncertainty of the experimental data.

There are plenty of further improvements available to correct the RPA.[13,54,64−68] Alternatively, there is the coupled cluster (CC) method as well. In each case, the RPA acts as a reasonably sound starting point, since it is equivalent to the direct ring coupled cluster approach and thus poses as a proper subset of the CC singles and doubles (CCSD).[69] Recently published methodological advancements[70,71] allow for tuned, regional CC corrections to the RPA correlation energy. This paves the way for even more accurate benchmarks that could include oxygen defect formation or heterogeneous catalysis at the ceria surface.

In conclusion, we present a novel benchmark of many-electron theory on strongly correlated materials, using ceria as a specific test case. We demonstrate that MP2 and RPA overcome the limits of DFT+$U$ and hybrid functionals in terms of parameter bias while providing robust geometries and energies. The RPA in particular provides unprecedented agreement with experimental values, outperforming all DFT flavors. Moreover, it forms the basis for additional corrections that systematically improve the accuracy. Here, we demonstrate that RPA+rSOX resides reliably within the experimentally measured results. Such enhanced accuracy likewise raises the bar on computational reference data. Benchmarking aside, the same methods also have potential applications in machine-learned density functionals.

## ASSOCIATED CONTENT

### Supporting Information

The Supporting Information is available free of charge at https://pubs.acs.org/doi/10.1021/acs.jpclett.1c01589.

> Computational settings and pseudopotentials, (anti)-ferromagnetic ground states and metastable states of solid $Ce_2O_3$, impact of the electronic structure on the $U$ dependence, theory behind the correlation methods, basis set convergence and thermodynamic limit of the reaction energies and lattice parameters, timings, treatment of isolated molecules in a periodic code, tables of reaction and atomization energies, and the experimental references (PDF)

## AUTHOR INFORMATION

### Corresponding Authors

**Tobias Schäfer** − *Institute for Theoretical Physics, TU Wien, 1040 Vienna, Austria*; orcid.org/0000-0003-0716-6516; Email: tobias.schaefer@tuwien.ac.at

**Nathan Daelman** − *Institute of Chemical Research of Catalonia, The Barcelona Institute of Science and Technology, 43007 Tarragona, Spain*; Email: ndaelman@iciq.es

### Author

**Núria López** − *Institute of Chemical Research of Catalonia, The Barcelona Institute of Science and Technology, 43007 Tarragona, Spain*; orcid.org/0000-0001-9150-5941

Complete contact information is available at:
https://pubs.acs.org/10.1021/acs.jpclett.1c01589

### Notes

The authors declare no competing financial interest.
VASP input and output files pertaining to the phase transition have been made available on the ioChem-BD database, ref 72.

## ACKNOWLEDGMENTS

This research has been supported by the Postdoctoral Orientation Program of the Ministerio de Economía y Competitividad (BES-2016-076361). The authors acknowledge BSC-RES and BIFI for generously providing access to their computational resources. The authors also acknowledge the TU Wien Bibliothek for financial support through its Open Access Funding Program. We also thank Georg Kresse and Andreas Grüneis for critically reading the manuscript and providing feedback.